# Echinoderms Have Bilateral Tendencies

Chengcheng Ji[1]*[9], Liang Wu[1][9], Wenchan Zhao[1], Sishuo Wang[1], Jianhao Lv[2]

1 College of Biological Sciences, China Agricultural University, Beijing, China, 2 College of Science, China Agricultural University, Beijing, China

## Abstract

Echinoderms take many forms of symmetry. Pentameral symmetry is the major form and the other forms are derived from it. However, the ancestors of echinoderms, which originated from Cambrian period, were believed to be *bilaterians*. Echinoderm larvae are bilateral during their early development. During embryonic development of starfish and sea urchins, the position and the developmental sequence of each arm are fixed, implying an auxological anterior/posterior axis. Starfish also possess the Hox gene cluster, which controls symmetrical development. Overall, echinoderms are thought to have a bilateral developmental mechanism and process. In this article, we focused on adult starfish behaviors to corroborate its bilateral tendency. We weighed their central disk and each arm to measure the position of the center of gravity. We then studied their turning-over behavior, crawling behavior and fleeing behavior statistically to obtain the center of frequency of each behavior. By joining the center of gravity and each center of frequency, we obtained three behavioral symmetric planes. These behavioral bilateral tendencies might be related to the A/P axis during the embryonic development of the starfish. It is very likely that the adult starfish is, to some extent, *bilaterian* because it displays some bilateral propensity and has a definite behavioral symmetric plane. The remainder of bilateral symmetry may have benefited echinoderms during their evolution from the Cambrian period to the present.

**Funding:** This work was supported by the grant from the National Natural Science Foundation of China (Project J1103520) and National Basic Science Foundation of China, (Education of Field Research, Project 0830630). The funders had no role in study design, data collection and analysis, decision to publish, or preparation of the manuscript.

**Competing Interests:** The authors have declared that no competing interests exist.

* E-mail: jichengcheng@yahoo.com.cn

[9] These authors contributed equally to this work.

## Introduction

There are six classes of echinoderms: Crinoidea, Asteroidea, Ophiuroidea, Echinoidea, Holothurioidea and Concentricycloidea. All are radial, except for the holothurians [1], which evolved from a radial ancestor [2]. Among the actinomorphic echinoderms, most are pentameral animals [1,2], and the other symmetry forms are derived from pentameral symmetry [3–5].

The ancestors of echinoderms, which originated from the Cambrian era, were *deuterostomes* [6]. Because *deuterostomes* are all bilateral, we can infer that the ancestors of echinoderms were *bilaterians* [7,8]. To adapt to their benthonic habitat and planktonic habitat niches, echinoderms evolved from bilateral symmetry first to asymmetry, then to pentameral symmetry [9–11].

Echinoderm larvae are bilateral during early development [12,13]. They first change into asymmetry [14,15] then continue to change into pentameral symmetry [11,16]. Larvae have an A/P axis that is determined by the fixed developmental mode and the absolute sequence in the embryonic development process [16]. During this process, tissues and organs match the corresponding symmetry forms [16,17].

The Hox gene family is known to control symmetry development and formation of the body axis in *deuterostomes* [18–21]. So, it is possible that the Hox genes in echinoderms control the change from bilateral to radial symmetry [22–25].

Despite often being studied, it is still not clear whether adult starfish retain their bilateral symmetry mechanism to some extent. Jennings and Cole studied the locomotion and righting of the starfish in *Stylasterias forreri* [26,27]. Jones etc. studied the leading arm during motion in *Echinaster sepositus* [28]. Kjerschow-Agersborg studied the physiological anterior end in *Pycnopodia helianthodes* [29]. Thorpe studied the orientation of locomotion in echinoderm [30]. O'Donoghue studied the migration behaviors of certain starfish [31]. Rodenhouse studied the morphology and behavior in *Pteraster tesselatus* [32]. Smith studied the neural system and behavior of starfish [33].

To explore this question, we weighed starfish arms and central disks to determine their center of gravity and symmetric mechanism. We then counted the number of times that starfish used each arm and statistically calculated their behavioral symmetric plane. We concluded that starfish are slightly bilateral in behavior, and they are, to some extent, bilateral animals.

## Materials and Methods

### Materials

*Asterias amurensis* is a very common species of sea star in East Asian coastal areas. We designated the arm opposite to the madreporite as Arm 1, and the others follow clockwise successively in aboral view (Fig. 1). Our numbering system is different than the previous ones, shown in Table 1 [34].

### Weighing

All weights were measured on analytical balances that were accurate to at least 0.0001 g or 0.1 mg.

The position of the last pair of side pedicellaria at the base of each arm was designated as the cutting line (line L in Fig. 2). The arms of the dried starfish were cut off and weighed. Because some starfish may regenerate a new arm when it is broken, those with notably different arms were excluded. Due to the large number of starfish used in this experiment, minor differences resulting from inconspicuous regeneration can be regarded as insignificant.

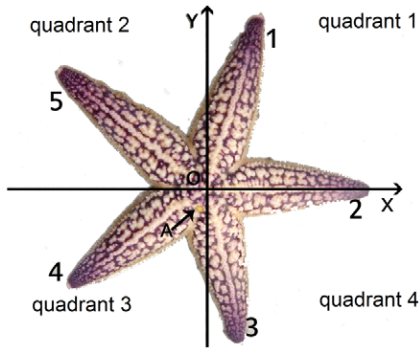

**Figure 1. The madreporite is shown by Point A.** Arm 1 is the arm opposite to the madreporite, and the other arms follow clockwise successively in aboral view. The coordinate system is as shown in the figure. Point O, the origin, is located at the center. Arm 2 lies on the positive x-axis.
doi:10.1371/journal.pone.0028978.g001

After cutting off all of the arms, the intact central disk was weighed (Fig. 2). Because the organs are within the central disk and the organs are soft with irregular shapes, it was difficult to divide the central disk into five equal parts and weigh them precisely.

### Behavioral experiment

The following experiments were conducted in calm seawater, and the starfish used were all healthy and sound. We lifted the starfish wholly, not just with one single arm. No specific permits were required for the described field studies.

**1) Turning-over experiment.** The starfish were turned upside down and left to turn back freely. Generally, the starfish firstly extended its arms upwards, then bent two adjacent arms against the ground for support, stamped the ground with the opposite arm and lifted the other two arms upwards on each side, finally the opposite arm lost its contact with the substrate and the starfish turned over [35,36] (Fig. 3). In such cases, we recorded the number of the stamping arm. Occasionally, they only bent one arm against the ground for support, stamped the ground with the opposite two arms and lifted the other two arms upwards on each side. In these cases, we recorded the two stamping arms and assigned each a weight of 0.5.

**2) Crawling experiment.** The starfish were placed in water and left to crawl freely. Generally, the starfish crawled with two arms forward as the leading arms, two on the side/rear and one backward [35,36] (Fig. 4). In such cases, we recorded the number of the backward arm. Occasionally, they crawled with one arm forward, two to the side/front and two backward. In such cases, we recorded the two arms backward and assigned each a weight of 0.5.

**3) Fleeing experiment.** The level of the seawater was lowered enough to bare the central disk of the starfish. A drop of alkali solution was placed in the center of the starfish's back, and we observed its escape. The same arms were recorded as in the crawling experiment.

### Statistical Methods

**1) Treatment of mass during statistics process.** During the statistics of mass, we compare the five arms of the starfish within itself, which means each starfish was designated with the same weight 5 and contributed the same to the sum weight.

**2) V-test.** V-tests were carried out on the weight and behavior data of each arm of the starfish in order to detect the tendency of skewing on one direction.

The course is as follows: Assume the data (either weight or behaviors) for each arm was d1, d2, d3, d4, d5, and D means the sum of the five data. And $\theta_i$ means the intersection angle between the arm and the designated direction.

Then $v = \Sigma(d_i/D)\cos\theta_i$, and $u = v(2n)^{1/2}$. Then the u value can be compared with the critical value in the v-test chart to identify the significance level $\alpha$ [37,38].

### Results

The center of gravity in a bilateral animal is supposed to lie on the plane of bilateral symmetry. Because its behavior on both sides is also bilateral, the behavioral center of frequency should also lie on the symmetric plane. A straight line can be drawn joining the

**Table 1.** Numbering systems.

| | | Clockwise oral view | | | | | Clockwise aboral view | | | | |
|---|---|---|---|---|---|---|---|---|---|---|---|
| | | **BIVIUM** | | **TRIVIUM** | | | **TRIVIUM** | | | **BIVIUM** | |
| **Coe 1912; Schuchert 1915** | | | | | | | | | | | |
| Preyer 1886-7; Schuchert 1915 | | 1 | M* | 5 | 4 | 3 | 2 | 2 | 3 | 4 | 5 | M* | 1 |
| Ludwig 1899; Kjerschow-Agersborg 1922; Rodenhouse & Guberlet 1946 | | I | M* | V | IV | III | II | II | III | IV | V | M* | I |
| Jennings 1907; Cole 1913 | | a | M* | e | d | c | b | b | c | d | e | M* | a |
| Polls & Goner 1975; Zirpolo 1928 | | A | M* | E | D | C | B | B | C | D | E | M* | A |
| Gemmill 1914; Chadwick 1923; Reid 1950; Smith 1950 | | I | M* | II | III | IV | V | V | IV | III | II | M* | I |
| Knight-Jones (pers. comm); Jones et al. 1968 | | 1 | M* | 2 | 3 | 4 | 5 | 5 | 4 | 3 | 2 | M* | 1 |
| Lovén 1874 Delage & Hérouard 1903 | | II | M* | III | IV | V | I | I | V | IV | III | M* | II |
| Bather 1900; Cuénot 1912; Hyman 1955 | | C | M* | D | E | A | B | B | A | E | D | M* | C |
| Moore & Fell 1966 Hotchkiss 1979, 1995, 1998 | | D | M* | E | A | B | C | C | B | A | E | M* | D |
| O'Donoghue 1926 | | R.1 | M* | A | L.1 | L.2 | R.2 | R2 | L2 | L1 | A | M* | R.1 |
| Numbering system in this article | | 4 | M* | 3 | 2 | 1 | 5 | 5 | 1 | 2 | 3 | M* | 4 |

M* means madreporite.
doi:10.1371/journal.pone.0028978.t001

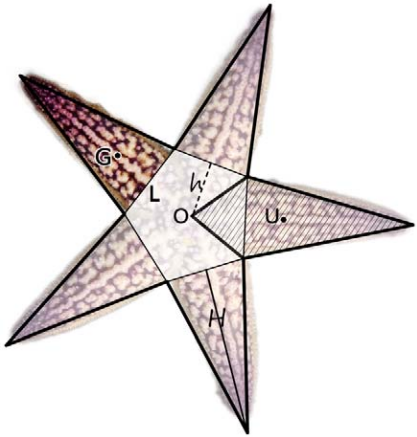

**Figure 2. The five-pointed star conforms to the real starfish's shape.** Line h, the distance between the center and bottom of the arm, is 1. Line H, the length of an arm, is 2.985. Line L is the cutting line. When weighing, the arms were cut off along Line L. The central disk is the light-colored regular pentagon. We considered the arms as cones, so Point G, the center of gravity of an arm, lay at H/4 of the triangle. We assumed the central disk to be homogeneous, so Point O, the center of gravity of the central disk, lay in the center of the regular pentagon. The unit of action is the shaded part. A unit of action contains one arm and 1/5 of the central disk near the arm. Point U is the center of gravity in the plane unit, and we assigned the frequency of action of each arm to it. Point U is located at $U = [(h \times h \times 2/3) + H \times (h + H/3)]/(h+H) = 1.6617$.
doi:10.1371/journal.pone.0028978.g002

two centers and is considered to be the vertical projection of the plane of bilateral symmetry. We obtained three planes from our three behavioral experiments.

We projected the starfish and the three planes of bilateral symmetry to the same plane and got the two-dimensional figure (Fig. 1). All computation was in a plane coordinate system with the projection of the planes of bilateral symmetry.

### Center of gravity

During calculation, we assumed that the central disk was homogeneous. The central disk/arm mass ratio of 1.43 is obtained from averaging the central disk/arm mass ratio of 50 starfish. We found that the ratio had little influence on the final calculation of the position of the center of gravity of the intact starfish. The maximum value 1.82 and the minimum 1.28 were tested and no significant changes were found. Thus, if we designate the mass of one arm as 1, the sum weight of an intact starfish is (5+1.43).

Suppose a starfish is a five-pointed star and the distance between the center and bottom of an arm is 1, which is h. Through software simulation, we adjusted the ratio until the five-pointed star figure aligned with the real starfish, ultimately obtaining the distance from the endpoint of arms and the center as 3.985. Thus the length of an arm H was 2.985.

We considered the arms to be cones, so the center of gravity of an arm is G, located at (H/4) near to the bottom of the triangle. This is to say that the distance from the origin to the point G is $(h+H/4) = 1.746$ (Fig. 2).

According to our calculation using data from 649 starfish (Table 2), the center of gravity of an intact starfish lies at the coordinates, $0.0044, -2.2746 \times 10^{-4}$, in the fourth quadrant. The center or gravity is considered to be almost the same to the origin point, v-test, $\alpha > 0.25$. And the weights of each arms were significantly different, tested with analysis of variance (ANOVA), $F = 26.35, \alpha < 0.05$.

### Center of frequency and symmetric planes

We calculated the center of frequency with frequentness of action in a plane analytic fashion. We assumed that a starfish is a planar, five-pointed star, with one arm and the 1/5 of the central disk near the arm as a unit of action. We then assigned the frequency of action of each arm to the centroid of the unit, shown as the point U in Fig. 2, $U = [(h \times h \times 2/3) + H \times (h + H/3)]/(h+H) = 1.6617$. The center of frequency was then obtained from averaging the centers of the five units. If the distance between the center and the bottom of an arm is equal to 1 and the distance between the center and the endpoint of an arm is equal to 3.985 (Fig. 2), then the line drawn between the center of frequency and the center of gravity is on the symmetric plane of the frequency of action (Fig. 5).

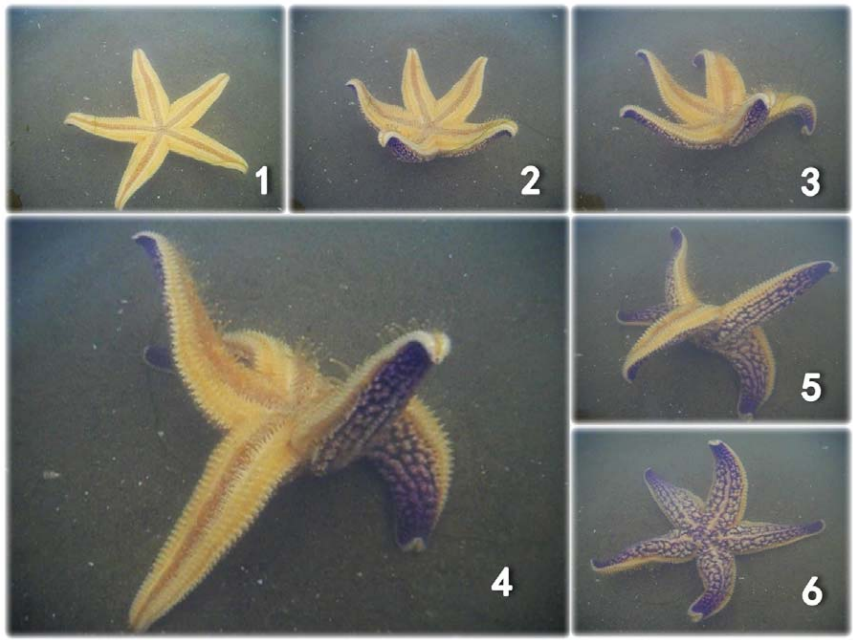

**Figure 3. The turning-over process is shown from Step 1 to 6.** Generally, the starfish firstly extended its arms upwards, then bent two adjacent arms against the ground for support, stamped the ground with the opposite arm and lifted the other two arms upwards on each side, finally the opposite arm lost its contact with the substrate and the starfish turned over.
doi:10.1371/journal.pone.0028978.g003

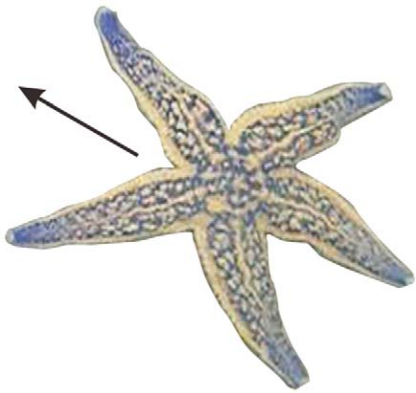

**Figure 4. Crawling action of the starfish.** Generally, the starfish crawled with two arms forward as the leading arms, two on the side/rear and one backward. The arrow indicates the direction of movement.
doi:10.1371/journal.pone.0028978.g004

### 1) Turning-over experiment

Table 3 shows the experimental data from 1,034 starfish.

The center of frequency in the turning-over experiment was located at the coordinates 0.0803, −0.0265, in the fourth quadrant. The center of frequency is not the same with the origin point, v-test, $0.01 < \alpha < 0.025$. The symmetric plane is shown in blue in Fig. 5.

### 2) Crawling experiment

Table 4 shows the experimental data from 694 starfish.

The center of frequency in the crawling experiment was located at the coordinates 0.0222, −0.0656, in the fourth quadrant. The center of frequency is not the same with the origin point, v-test, $0.05 < \alpha < 0.1$. The symmetric plane is shown in yellow in Fig. 5.

### 3) Fleeing experiment

Table 5 shows the experimental data from 548 starfish.

The center of frequency in the fleeing experiment was located at the coordinates 0.0749, −0.0364, in the fourth quadrant. The center of frequency is not the same with the origin point, v-test, $0.025 < \alpha < 0.05$. The symmetric plane is shown in green in Fig. 5.

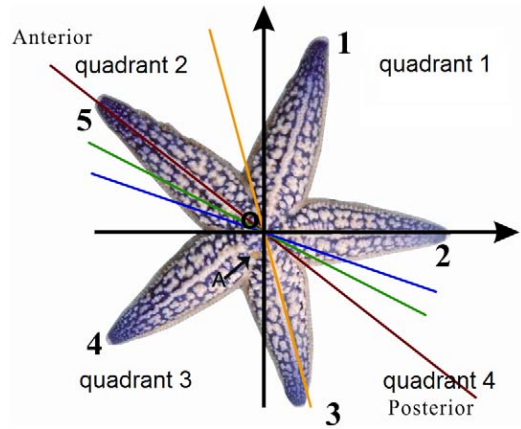

**Figure 5. The coordinate system is the same as in Figure 1.** The blue, yellow, green and red planes represent the symmetric planes of turning-over, crawling, fleeing and average, respectively. Anterior and posterior directions are as shown.
doi:10.1371/journal.pone.0028978.g005

Turning-over is a difficult action for a starfish. Crawling represents the propensity of the starfish to use its body when it is free moving. Fleeing represents the organism's emergency response when it is in danger. These three experiments show the symmetric plane in three ways, so the three centers of frequency can be averaged: 0.0591, −0.0428 in the fourth quadrant. The average plane is shown in red in Fig. 5.

V-tests were carried out on the frequency data of the turning-over, crawling and fleeing action, to the direction of the average symmetric plane, $\alpha < 0.025$, $\alpha < 0.1$, $\alpha < 0.05$, respectively.

Fleeing occurs when a starfish is in danger, determining whether it will survive. The starfish must use its body to move quickly, and its fleeing behavior strongly reflects its bilateral propensity. Turning-over is difficult to execute, and the starfish must also efficiently use its body. Therefore, this behavior also reflects its bilateral propensity quite clearly. Finally, crawling is performed when the starfish is freely moving, so this exhibits the lowest bilateral propensity.

The plane of bilateral symmetry should lie approximately where the red line is (i.e., from Arm 5 to between Arms 2 and 3) (Fig. 5). We presume the A/P axis in the embryonic development might be relevant to the bilaterally symmetric behaviors in the adults. Thus we hypothesize that during movement adult starfish tend to take as front the direction of the Anterior end in the embryonic development, and take as rear the Posterior end. Also, there might to some extent exist some remaining of the A/P axis in the

**Table 2.** Relative weight of each arm (Standard Deviation).

| Arm order | 1 | 2 | 3 | 4 | 5 |
|---|---|---|---|---|---|
| Sum of the Relative weight | 656.0942 | 648.2409 | 652.6217 | 647.2929 | 640.7502 |
| Averaged Relative weight | 1.0109 (0.0435) | 0.9988 (0.0431) | 1.0056 (0.0462) | 0.9974 (0.0458) | 0.9873 (0.0433) |

Analysis of Variance: $F = 26.35, \alpha < 0.05$.
v-test: $\alpha > 0.25$.
doi:10.1371/journal.pone.0028978.t002

**Table 3.** Relative turning-over frequency.

| Arm order | 1 | 2 | 3 | 4 | 5 | Standard Deviation |
|---|---|---|---|---|---|---|
| Number of times each arm was recorded | 205.5 | 229.5 | 216 | 197 | 186 | - |
| Proportion of the number recorded | 0.1987 | 0.2220 | 0.2089 | 0.1905 | 0.1799 | 0.0145 |

v-test: $0.01 < \alpha < 0.025$.
doi:10.1371/journal.pone.0028978.t003

Table 4. Relative crawling frequency.

| Arm order | 1 | 2 | 3 | 4 | 5 | Standard Deviation |
|---|---|---|---|---|---|---|
| Number of times each arm was recorded | 131 | 148 | 140 | 153.5 | 121.5 | - |
| Proportion of the number recorded | 0.1888 | 0.2133 | 0.2017 | 0.2212 | 0.1751 | 0.0168 |

v-test: $0.05<\alpha<0.1$.
doi:10.1371/journal.pone.0028978.t004

Table 5. Relative fleeing frequency.

| Arm order | 1 | 2 | 3 | 4 | 5 | Standard Deviation |
|---|---|---|---|---|---|---|
| Number of times each arm was recorded | 127 | 128 | 112 | 131.5 | 85.5 | - |
| Proportion of the number recorded | 0.2175 | 0.2192 | 0.1918 | 0.2252 | 0.1464 | 0.0291 |

v-test: $0.025<\alpha<0.05$.
doi:10.1371/journal.pone.0028978.t005

adults' body [3,39]. Even though we know from the statistically studied behavior that starfish are *bilaterians*, the adult starfish only retains a slight mechanism of bilateral symmetry, and we can only roughly identify the location of the symmetric plane.

As with the orientation of the starfish's movement, Arm 5 obviously moves much less in the backward direction (i.e., starfish tend to move toward the direction where Arm 5 is located). The direction of Arm 5 can, therefore, be considered anterior, and the direction in between Arm 2 and Arm 3 can be considered posterior (Fig. 5).

## Discussion

From our behavioral research, we can conclude that starfish behave as *bilaterians*. Our findings can be generalized to all classes of echinoderms except for the sea cucumber. In other words, during their evolution from the Cambrian era to present, some bilateral symmetry has persisted in adult echinoderms. It is also likely that other systems match this bilateral symmetry, such as the nervous system, the sensory organs and the motor system.

Animals tend to move in the anterior direction, instead of posterior or other directions. Sense organs and the nervous system also tend to concentrate at the anterior side. Because starfish are more likely to move toward Arm 5, it is possible that their sensory organs are focused there as well. Nerve ganglions may also be more developed around this direction, and the nervous system in the central disk might concentrate towards this direction.

The essence of the radial symmetry of echinoderms is an adaption to their benthic habitat niche. The sensory organs of starfish are not highly developed, nor can they move very fast. Starfish tend to act more like radial animals so that they can be open to stimuli from all directions and move evenly toward different directions. But when they are in danger, they tend to use their bodily functions efficiently and behave more bilaterally. This remnant of their bilaterally symmetric ancestors may have benefited echinoderms during evolution. Concentration of their sensory organs and nervous system helps echinoderms observe and react to their environment more intensively, especially for tracing prey and detecting enemies. Concentration of their motor system helps to save energy and move faster. Additionally, their propensity for motion makes it more convenient to pursue prey and run away from natural enemies. Partial bilateral symmetry facilitates several difficult actions. Echinoderms resemble the octopus to some extent, appearing like a radial animal, while actually having bilateral behavioral mechanisms and the corresponding physical characteristics.

Our results provide evidence that echinoderms have retained bilateral tendencies from the Cambrian era to the present, and this likely has some kind of adaptive significance. We have verified the existence of the A/P axis in auxology and its influence on starfish from birth to adulthood. This work has implications for research on the evolution, embryonic development, behavior and fossil research of echinoderms and *deuterostomes*.

## Acknowledgments

We wish to thank the 2 referees for their helpful suggestions for improving the manuscript.

## Author Contributions